\documentclass[12pt]{article}
\usepackage{color,amsmath,latexsym,amssymb,amsthm}
\usepackage{enumitem}
\usepackage{amsmath}
\usepackage{graphicx}
\usepackage{textcomp}
\def\and{\par\vskip 2.25em minus 1em\moveright\leftskip\vbox{\hrule width
\leftskip}}

\def\title{A Novel Convolution-Based Stratified Attribute Estimator for QRE Determination in R\&D Tax Credit Studies}
\def\author{Deborah Lynn Goldwasser, PhD}
\def\affiliation{Florida International University}
\def\email{dgoldwas@fiu.edu}
\def\keywords{Statistical Sampling, R\&D Tax Credit, Qualified Research Expenses, Internal Revenue Service}
\def\donemonth{September 18,}
\def\doneyear{2026}
\def\department{Department of Mathematics and Statistics}
\def\address{11200 SW 8th Street, Miami, FL 33199}

\begin{document}

%----------title page--------------------------------
% the title page is automatically generated using the definitions
%   provided in the file titledef.tex
% These should not requre editing as they areset to the thesis
%   guidelines, however small spacing adjustments may be necessary
%   depending on the size of your committee, and/or the length of your
%   title
\begin{titlepage}
\centering

% Title
{\Large \textbf{\title} \par}

\vspace{1.5em}

% Authors
{\large \author \par}

\vspace{1em}

% Affiliations
{\small \affiliation \par}

\vspace{1em}

{\small \department \par}

\vspace{1em}

{\small \address \par}

\vspace{1em}

% Corresponding author
{\small \textbf{Corresponding author:} \email \par}

\vspace{2em}

% Abstract
\begin{minipage}{0.9\textwidth}
\small
\textbf{Abstract} \\
The One Big, Beautiful Bill Act reinstates immediate expensing of domestic research expenses, reducing the tax burden and incentivizing reinvestment in U.S.-based research and development including software development across a wide range of industries. In this context, accurate and defensible estimation of qualified research expenses (QREs) is of high importance. Statistical sampling provides a practical framework for estimating QREs for a well-defined population of business components (sampling frame) documented in accordance with Internal Revenue Code Section 41 (Form 6765). IRS Revenue Procedure 2011-42 permits both attribute and variable statistical methods, although the latter (stratified mean and difference estimators) are often regarded as the standard approach to QRE estimation. \\\\
In this paper, we compare attribute and variable statistical methods within a simulation study and demonstrate that attribute methods offer several theoretical and practical advantages. A key concern with the simple attribute estimator is the possibility of upward bias in QRE determination arising from a preponderance of high potential QRE (pQRE), non-qualified projects in the sampling frame. We address this concern by introducing a stratified sampling design that ensures adequate representation of high pQRE projects in the sample. We demonstrate that a convolution-based stratified attribute estimator produces a valid one-sided 95\% lower confidence bound on total QREs across a range of sampling frame structures.

\end{minipage}

\vspace{1.5em}

% Keywords
\begin{minipage}{0.9\textwidth}
\small
\textbf{Keywords:} \keywords
\end{minipage}

\vfill

% Date
{\small \donemonth\ \doneyear \par}

\end{titlepage}

%-----general adjustments-----------------------------
\pagestyle{myheadings}
\renewcommand{\baselinestretch}{1.5} % spacing between lines, set to
\small\normalsize                    % '1.5' or '2' for final version
                                     % use '1' for single-spaced drafts

%\pagenumbering{empty}  % lower case roman numerals for front matter
%\setcounter{page}{2}   % page 1 was the title page, start with page 2

%-----personal text adjustments
% I used the following commands to help adjust a little of the final
%   formatting for tables and paragraphs.  They are not required for
%   the thesis format, but if you have large tables I would at least
%   recommend using this first command.  Uncomment them if you wish.
%
%\renewcommand{\arraystretch}{0.75} % reduce spacing in tables back
%                                   % to something reasonable
%                                   % (single-spaced)
%
%\setlength{\parindent}{0em} % get rid of paragraph indentation,
%\setlength{\parskip}{2ex}   % instead place a blank line between
%                            % paragraphs

%-----body-------------------------------------------------
\clearpage \pagestyle{empty}
%\begin{figure}
  %\centering
  %Requires \usepackage{graphicx}
  %\includegraphics[width=]{}\\
  %\caption{}\label{}\\
%\end{figure}

%%%%%%%%%%%%%%%%%%%%%%%%%%%%%%%%%%%%%%%%%%%%%%%%%%%%%%%%%%
% section 1
%%%%%%%%%%%%%%%%%%%%%%%%%%%%%%%%%%%%%%%%%%%%%%%%%%%%%%%%%%
\section*{\text {Introduction}}
\subsection*{\text {Background and Motivation}}
On July 4, 2025, the One Big Beautiful Bill Act (OBBBA) was signed into law, restoring immediate expensing of domestic research expenses, thereby incentivizing American companies to invest in domestic research and development (R\&D) including software development\textsuperscript{1}. A wide range of American industries can benefit from this tax reduction, not limited to technology, engineering and scientific industries. American companies innovating in farming methods and producing novel architectural designs can also benefit\textsuperscript{2,3}. The tax relief provided in the OBBBA is especially important to small and medium-sized businesses who need access to liquid capital to grow.  Despite these significant benefits, accessing the credit can be challenging. The Internal Revenue Service (IRS) documentation requirements are rigorous, requiring businesses to clearly substantiate their qualifying activities and expenses\textsuperscript{4}. Under Internal Revenue Code Section 41, eligibility for the credit is determined at the business component level, each of which is subject to a strict four-part test\textsuperscript{5}. 
\subsection*{\text {Role of Statistical Sampling}}
Research conducted within a firm can be broad in scope, often encompassing numerous business components and underlying projects where the exact qualified research expenses (QREs) are not readily known. Statistical sampling offers a practical and efficient framework for estimating QREs across a well-defined population of business components (sampling frame) or a related population that has a direct nexus to those components. By using a sampling approach, taxpayers can avoid examining every individual unit within the sampling frame, significantly improving the manageability of the R\&D tax credit estimation process.\\\\
The sampling approach involves the following steps: 
\begin{enumerate}
\item Clearly define the sampling frame and the underlying sample units (well-aligned with business components).  When feasible, assign potential QREs (pQREs) to each sample unit.
\item Select a random sample, often a stratified random sample. 
\item Thoroughly document QREs associated with each unit in the random sample. 
\item Apply appropriate statistical methods to project total QREs for the full population based on the sampled records.
\end{enumerate}
Designated IRS statistical sampling coordinators (SSCs) review the sampling design and statistical calculations to ensure the validity of results\textsuperscript{6}.\\\\
IRS Revenue Procedure 2011-42 provides detailed guidance on the use of statistical sampling and estimation techniques for tax purposes\textsuperscript{7}. It permits variable methods which directly estimate QRE amounts (stratified mean and difference estimators) as well as attribute methods which estimate the proportion of qualifying units in the sampling frame and subsequently rescale total pQREs accordingly. Statistical methods must accurately quantify sampling variability in order to ensure that a QRE estimate is conservative for a typical random sample. IRS Revenue Procedure 2011-42 requires taxpayers to use a one-sided 95\% confidence bound that is least advantageous to the taxpayer.  However, a point estimate can be used in lieu of the one-sided 95\% confidence bound when sufficient precision can be demonstrated (i.e. relative precision $<$ 10\%).   
\subsection*{\text {IRS Scrutiny and Emerging Tax Court Guidance}}
The use of statistical sampling introduces inherent variability to QRE estimation, as different samples drawn from the same sampling frame can produce different QRE estimates. Different statistical methods applied to the same random sample can also produce different QRE estimates.  Furthermore, the validity of a sample-based QRE projection hinges strongly on the structure of the sampling frame and the underlying business components. Accordingly, sample-based QRE estimates are subject to heightened IRS scrutiny and United States Tax Court (Tax Court) review. In instances of a contested research credit, the taxpayer has the burden of proof in establishing the credit is valid\textsuperscript{8}.
\subsubsection*{\text {Selection of a "Representative" Sample}}
A primary concern of several Tax Court cases is the selection of a "representative" sample of projects to present at trial in order to determine the research credit in its entirety. In \emph{Suder v. Commissioner} (2014), both parties agreed pre-trial upon a binding sample of 12 out of 76 projects\textsuperscript{9}. The Tax Court upheld the research credit based on the results of the four-part test for that agreed-upon sample.  In \emph{Kapur v. Commissioner} (2024), by contrast, the parties could not agree upon a binding sample\textsuperscript{10}. The Tax Court rejected the Petitioner's proposal to limit discovery to just two large projects reflecting 72\% of pQREs and disallowed the claimed QREs. \\\\
The premise that a “representative” random sample can be reliably identified \emph{prima facie} is problematic from a statistical standpoint.  Just as a coin flip may result in 9 heads out of 10 tosses, a random sample may identify a disproportionate number of qualified projects relative to the sampling frame.  Unlike with the coin flip, where 50\% of tosses are expected to turn up heads, the true nature of the sampling frame is unknown \emph{a priori}. The purpose of statistical inference is to account for this sampling variability and limit overestimation of QREs based on "non-representative" samples to under 5\% of all samples. Nonetheless, larger samples are certainly desirable in that they increase the precision of the final QRE estimate.  Variable statistical methods also rely on large samples to satisfy asymptotic conditions necessary for valid statistical inference. 
\subsubsection*{\text {Defining the Sampling Frame}}
Another key consideration is the extent to which the sampling frame itself should be subject to discovery\textsuperscript{8}.  In \emph{Kapur v. Commissioner} (2024), the Respondents argued that in order to determine whether any sample is representative, basic structural information about the sampling frame must be obtained\textsuperscript{10}.  The Tax Court, while disallowing the particular research credit, did affirm, however, that it has the authority to limit discovery to a sample, when appropriate. In \emph{Bayer Corp. \& Subsidiaries v. United States} (2012), the district court ruling emphasized that the entire sampling frame must be well-defined in order to produce a valid statistical sample\textsuperscript{11}. \\\\ 
Were the courts to impose extensive discovery on the entire sampling frame, it would negate the cost and time-saving benefits of a statistical sampling approach.  Nonetheless, sound sampling theory implicitly relies on having a well-defined population from which to draw a random sample\textsuperscript{12}.  Furthermore, "haircut"-based estimation approaches such as the stratified difference estimator and attribute methods rely on a known amount of pQREs for the entire sampling frame. The "haircut" amount is determined based on the sample results and pQREs are then reduced accordingly.  The expectation that these pQREs can be substantiated is certainly reasonable and appropriate.\\\\ 
Recently enhanced documentation requirements reflect a renewed emphasis on having a well-defined sampling frame at the business component level.  For research expenses incurred in 2026 and later, Form 6765 must explicitly list up to 50 of the top business components associated with R\&D expenses\textsuperscript{13}.  Nonetheless, Form 6765 explicitly permits the use of statistical methods, preserving the ability to rely on sampling while ensuring that the underlying sampling frame is sufficiently documented.  
\subsubsection*{\text {Selection of a Statistical Method}}
The Tax Court has naturally also considered the specific statistical methods utilized to project total QREs from the random sample. IRS Revenue Procedure 2011-42 does not declare a preference for either attribute and variable methods, nor is it specific to QRE estimation only.  Even so, the IRS has taken the position that since variable methods utilize a direct nexus between QREs and individual projects, they are viewed more favorably as a direct means of QRE estimation. Several recent cases and expert witness testimony have addressed this topic, notably \emph{Max v. Commissioner} (2021), \emph{Felker v. Commissioner} (Docket No. 3871-17), and \emph{Intermountain Electric Inc. v. Commissioner} (Docket No. 11019-19) although no Tax Court ruling has provided definitive guidance on the selection of statistical methods as of the time of this publication\textsuperscript{14,15,16}.
\subsection*{\text {Goals of Paper}}
Given the higher levels of scrutiny that statistical sampling approaches face, this paper aims to clarify, demystify and highlight the relative advantages and weaknesses of the prevailing estimation procedures permitted by IRS Revenue Procedure 2011-42. We evaluate the performance of attribute and variable statistical methods under different sampling frame structures within a simulation study. We subsequently propose a novel convolution-based stratified attribute estimator not currently described by IRS Revenue Procedure 2011-42 and demonstrate its advantages. In addition, we find that a simple stratified attribute estimator performs reasonably well by ensuring adequate representation of projects having the highest pQREs in the sampling frame.
\section*{\text {Methods}}
\subsection*{Model Framework}
Given a sampling frame of \emph{N} potentially qualified R\&D projects divided into \emph{k} strata (i.e. $N = \sum_{i=1}^{k} N_i$), let T denote total pQREs such that $T = \sum_{i=1}^{k} T_{i}$. Furthermore, let $T_{Q,i}$ and $T_{NQ,i}$ equal the total pQREs associated with qualified and non-qualified R\&D projects, respectively, in the $i^{\text{th}}$ stratum, whereby $T_{i} = T_{Q,i} + T_{NQ,i}$. Therefore, $T_Q$, actual QREs in the sampling frame, equals $\sum_{i=1}^{k} T_{Q,i}$ since $T_{Q,i}$ reflects actual QREs in the $i^{\text{th}}$ stratum.\\\\
In order to draw a mathematical comparison between attribute and variable methodologies, we assume that projects are either wholly qualified or wholly non-qualified. Let $N_{Q,i}$ and $N_{NQ,i}$ denote the total number of qualified and non-qualified R\&D projects, respectively, in the $i^{\text{th}}$ stratum, so that $N_i = N_{Q,i} + N_{NQ,i}$. The proportion of qualified R\&D projects in the $i^{\text{th}}$ strata equals $p_i = \frac{N_{Q,i}}{N_i}$. 
\subsubsection*{Definitions of Key Parameters}
Define the following distinct mean parameter terms as follows:
\begin{itemize}
\item $\mu_{d,i} = \frac{-T_{NQ,i}}{N_i}$ where $\mu_{d,i}$ equals the average difference between actual QREs and pQREs per project in the $i^{\text{th}}$ stratum.\\  
\item $\mu_{i} = \frac{T_{Q,i}}{N_i}$ where $\mu_{i}$ equals the average actual QREs per project in the $i^{\text{th}}$ stratum.\\ 
\item $\mu_{Q,i} = \frac{T_{Q,i}}{N_{Q,i}}$ where $\mu_{Q,i}$ equals the average actual QREs per qualified R\&D project in the $i^{\text{th}}$ stratum.\\  
\item $\mu_{NQ,i} = \frac{T_{NQ,i}}{N_{NQ,i}}$ where $\mu_{NQ,i}$ equals the average non-qualified pQREs per non-qualified R\&D project in the $i^{\text{th}}$ stratum.\\
\item $\mu_{T,i} = \frac{T_i}{N_i}$ where $\mu_{T,i}$ equals the average pQREs per project in the $i^{\text{th}}$ stratum.
\end{itemize}
\subsection*{\text {Parameter Estimation Framework}}
With these key parameter definitions established, Table 1 illustrates how different statistical methods arrive at a determination of $T_Q$, total actual QREs for the sampling frame either with or without stratification.  Without stratification, the sampling frame has just one strata, the sampling frame in its entirety and the index \emph{i} can be omitted from the above definitions. 
\begin{table}[ht]
\centering
\caption{Determination of Actual QREs ($T_Q$) by Different Statistical Methods}
\label{tab:parameter_definitions}
\begin{tabular}{|l|c|c|}
\hline
\textbf{Statistical Method} &
\textbf{w/o Stratification} &
\textbf{w/ Stratification} \\
\hline
Attribute &
$T_Q = pT$ &
$T_Q = \sum_{i=1}^{k} p_i T_i$ \\
\hline
Variable (Mean Estimator) &
$T_Q = \mu N$ &
$T_Q = \sum_{i=1}^{k} \mu_i N_i$ \\
\hline
Variable (Difference Estimator) &
$T_Q = T + \mu_d N$ &
$T_Q = T + \sum_{i=1}^{k} \mu_{d,i} N_i$ \\
\hline
\end{tabular}
\end{table}
\subsubsection*{\text {Attribute Estimation of $T_Q$ Without Stratification}}
The objective of attribute statistical methods is to estimate the parameter p, the proportion of qualified R\&D projects in the sampling frame.  Multiplication of total pQREs (T) by p determines $T_Q$.  This approach produces a "haircut" equal to $(1-p)T$ that rescales total pQREs, since $pT = T - (1-p)T$.  Furthermore, the "haircut" equal to $(1-p)T$ is algebraically equivalent to $N_{NQ} \times {\mu}_T$. Consider the relationship between $T_Q$ and pT under different sampling frame structures.
\subsubsection*{\textbf{Scenario 1: $\mu_{Q} \approx \mu_{NQ} \approx \mu_{T}$}.}
When this condition holds, it follows that $p = \frac{N_Q}{N}=\frac{\mu_{T}N_Q}{\mu_{T}N} \approx \frac{\mu_{Q}N_Q}{\mu_{T}N} = \frac{T_Q}{T}$ and therefore $T_Q \approx pT$. In this case, the rescaled pQREs, pT, approximately equal $T_Q$, actual QREs associated with qualified R\&D projects in the sampling frame.
\subsubsection*{\textbf{Scenario 2: $\mu_{Q} > \mu_{NQ}$ and $\mu_{Q} > \mu_{T}$}.}
When this condition holds, it follows that $p = \frac{N_Q}{N}=\frac{\mu_{T}N_Q}{\mu_{T}N} < \frac{\mu_{Q}N_Q}{\mu_{T}N} = \frac{T_Q}{T}$ and therefore $T_Q > pT$. In this case, the rescaled pQREs, pT, are less than $T_Q$, and thereby underreport actual QREs associated with qualified R\&D projects in the sampling frame.
\subsubsection*{\textbf{Scenario 3: $\mu_{Q} < \mu_{NQ}$ and $\mu_{Q} < \mu_{T}$}.}
When this condition holds, it follows that $p = \frac{N_Q}{N}=\frac{\mu_{T}N_Q}{\mu_{T}N} > \frac{\mu_{Q}N_Q}{\mu_{T}N} = \frac{T_Q}{T}$ and therefore $T_Q < pT$. In this case, the rescaled pQREs, pT, are greater than $T_Q$, and thereby overreport actual QREs associated with qualified R\&D projects in the sampling frame.  
\subsubsection*{\text {Attribute Estimation of $T_Q$ With Stratification}}
When the sampling frame is stratified into \emph{i} strata, $\sum_{i} p_i \times T_i$ determines $T_Q$. This approach produces a "haircut" equal to $\sum_{i} (1-p_i) \times T_i$ that rescales total pQREs (T), since $\sum_{i} p_i \times T_i = T - \sum_{i} (1-p_i) \times T_i$. The relationship between $T_Q$ and $\sum_{i} p_i \times T_i$ depends on the overall balance of $\mu_{Q,i}$, $\mu_{NQ,i}$ and $\mu_{T,i}$ across the \emph{i} strata.
\subsubsection*{\text {Variable Estimation of $T_Q$}}
\textbf{\underline{Stratified Mean Estimator}}\\\\
The objective of the stratified mean estimator is to estimate $\mu_i$ within each of the \emph{i} strata. $T_Q$ is determined by $T_Q = \sum_{i} T_{Q,i} = \sum_{i} \mu_i \times N_i$. \\\\
\textbf{\underline{Stratified Difference Estimator}}\\\\
The objective of the stratified difference estimator is to estimate $\mu_{d,i}$ within each of the \emph{i} strata. $T_{Q}$ is determined by $T_Q = T - T_{NQ} = T - \sum_{i} T_{NQ,i} = T + \sum_{i} \mu_{d,i} \times N_i$, whereby the "haircut" on total pQREs (T) equals -$\sum_{i} \mu_{d,i} \times N_i$. 
\subsection*{Statistical Estimation of $T_Q$}
\subsubsection*{Simple Attribute Method: Without Stratification}
For a sampling frame of size \emph{N} and a random sample of size \emph{n}, determine \emph{e}, the number of erroneous, non-qualified projects in the sample. A point estimate for p is $\hat{p} = \frac{n-e}{n}$ and a point estimate for $T_Q$ is $\hat{T_Q} = \hat{p}T$.\\\\
Define the random variable \emph{E} as the number of erroneous, non-qualified projects in a random sample.  \emph{E} follows a hypergeometric probability distribution with parameters \emph{N}, \emph{n} and $N_{NQ}$, as follows:
\begin{center}
$P(E = e) = \frac{ \binom{N_{NQ}}{e}\binom{N-N_{NQ}}{n-e}}{\binom{N}{n}}$
\end{center}
Given a fixed value of \emph{e}, a one-sided 5\% lower confidence bound (LCB) on the total number of non-qualified projects ($N_{NQ}$) is the smallest value of z satisfying the following:
\begin{center}
$P(E \leq e | N_{NQ}) \leq 0.05$ for all $N_{NQ} \geq$ z
\end{center}  
The one-sided 5\% LCB \emph{z} serves as a lower bound on the rejection region of $N_{NQ}$, total non-qualified R\&D projects in the sampling frame.  Define z* as the largest integer value strictly less than z when z is greater than zero and zero otherwise. The value z* reflects the one-sided 95\% upper confidence bound on $N_{NQ}$. A rescaling factor r defined as r = $\frac{N-z*}{N}$ reflects the one-sided 95\% LCB on p. Given total pQREs in the sampling frame (T), a simple attribute LCB on actual QREs equals $\text{LCB}_{SA}(\hat{T_Q}) = rT$. The "haircut" on pQREs produced by the $\text{LCB}_{SA}$ estimator equals z*$\times {\mu}_T$, where ${\mu}_T = \frac{T}{N}$. 
\subsubsection*{Stratified Attribute Methods}
The sampling frame of size \emph{N} potentially qualified R\&D projects is partitioned into \emph{k} strata based on pQREs. A stratified random sample of size \emph{n} is drawn where $n = \sum_{i=1}^{k} n_i$ and $n_i$ equals the sample size in the $i^{\text{th}}$ stratum.  Stratified random sampling ensures adequate representation of projects in upper strata with higher than average pQREs. Define \emph{$e_i$} as the number of erroneous, non-qualified projects in the  $i^{\text{th}}$ stratum where $e = \sum_{i=1}^{k} e_i$. A point estimate for $p_i$ is $\hat{p_i} = \frac{n_i-e_i}{n_i}$ and a point estimate for $T_Q$ is $\hat{T_Q} = \sum_{i=1}^{k} \hat{p_i} \times T_i$.\\\\
\textbf{\underline{Simple Stratified Attribute LCB}}\\\\
Given a fixed value of \emph{e}, the rescaling factor \emph{r} is computed as described above and multiplied by total pQREs (T) to produce a simple stratified attribute LCB on actual QREs: $\text{LCB}_{SSA}(\hat{T_Q}) = rT$. \\\\ 
\textbf{\underline{Convolution-Based Stratified Attribute LCB}}\\\\
Given a fixed value of \emph{e}, let the matrix M denote all possible allocations of less than or equal to \emph{e} erroneous, non-qualified projects across the \emph{k} strata.  The number of rows in matrix M equals $\sum_{j=0}^{e} \binom{j+k-1}{k-1} = \binom{e+k}{k}$ and the number of columns equals the number of strata \emph{k}.\\\\
Define the random variable {$E_i$} as the number of erroneous, non-qualified projects detected among the $n_i$ sampled projects in the $i^{\text{th}}$ stratum where $E = \sum_{i=1}^{k} E_{i}$. Define the k-dimensional parameter vector $\boldsymbol{\theta}$ equal to the number of non-qualified projects across the k strata where
\begin{center}
$\boldsymbol{\theta} = (N_{NQ,1},N_{NQ,2},...,N_{NQ,k})$
\end{center}
Conditional on a fixed $\boldsymbol{\theta}$, each \emph{$E_i$} follows a hyper-geometric probability distribution with parameters \emph{$N_i$}, \emph{$n_i$} and $N_{NQ,i}$, as follows:
\begin{center}
$P(E_i = e_i) = \frac{ \binom{N_{NQ,i}}{e_i}\binom{N_i-N_{NQ,i}}{n_i-e_i}}{\binom{N_i}{n_i}}$
\end{center}
The probability $P(E \leq e|\boldsymbol{\theta})$ can be computed using matrix M based on a convolution formula as follows:
\begin{center}
$P(E \leq e|\boldsymbol{\theta}) = \sum_{j=0}^{e} \sum_{m_{.1}+m_{.2}+...+m_{.k}=j} \prod_{i=1}^{k} P(E_i = m_{.i}|\boldsymbol{\theta})$
\end{center}
When $P(E \leq e|\boldsymbol{\theta})$ is less than 0.05 then the unique parameter vector $\boldsymbol{\theta}$ lies in the 5\% rejection region ($\mathcal{R}_{0.05}$) defined as follows:
\begin{center}
$\mathcal{R}_{0.05} = \{\boldsymbol{\theta}: P(E \leq e|\boldsymbol{\theta}) \leq 0.05\}$
\end{center}
Analogously, the 95\% acceptance region ($\mathcal{A}_{0.95}$) is defined as follows: 
\begin{center}
$\mathcal{A}_{0.95} = \{\boldsymbol{\theta}: P(E \leq e|\boldsymbol{\theta}) \geq 0.05\}$
\end{center}
For each $\boldsymbol{\theta}$ in $\mathcal{A}_{0.95}$, $T_Q (\boldsymbol{\theta})$ can be computed as follows:
\begin{center}
$T_Q (\boldsymbol{\theta}) = \sum_{i=1}^{k} \frac{N_i-\boldsymbol{\theta_{i}}}{N_i} \times T_i$ 
\end{center}
The convolution-based stratified attribute LCB on actual QREs ($\text{LCB}_{CSA}(\hat{T_Q})$) is then determined by minimizing $T_Q$ over $\mathcal{A}_{0.95}$ where:
\begin{center}
$\text{LCB}_{CSA}(\hat{T_Q}) = \min_{\boldsymbol{\theta} \in \mathcal{A}_{0.95}} T_Q (\boldsymbol{\theta})$
\end{center}
Determination of $\text{LCB}_{CSA}(\hat{T_Q})$ requires a search over the parameter space in $\mathcal{A}_{0.95}$.  Note that within the $i^{\text{th}}$ stratum, both $T_Q(\boldsymbol{\theta})$ and $P(E \leq e|\boldsymbol{\theta})$ monotonically decrease as $N_{NQ,i}$ increases, holding all other components of {$\boldsymbol{\theta}$} constant. A branch-and-bound search algorithm is therefore used to search $\mathcal{A}_{0.95}$ to determine $\text{LCB}_{CSA}(\hat{T_Q})$. 
\subsubsection*{Stratified Variable Methods}
\textbf{\underline{Stratified Mean Estimator}}\\\\
Given a stratified random sample, average QREs ($\bar{x}_i$) within the $i^{\text{th}}$ stratum are estimated by $\bar{x}_i = \frac{\sum_{j=1}^{n_i} x_{ij}}{n_i}$. In this notation,  $x_{ij}$ is the actual QRE amount for the $j^{\text{th}}$ sampled project in the $i^{\text{th}}$ stratum.  A point estimate of $T_Q$, actual QREs for the sampling frame, is $\hat{T_Q} = \hat{X}_{SM} = \sum_{i} N_{i} \bar{x}_i$.\\\\
The sample variance ($s_{x_i}^2$) within each stratum equals $s_{x_i}^2 = \frac{ \sum_{j=1}^{n_i} x_{ij}^2 - n_i \bar{x}_i^2}{n_i - 1} $. The standard error of the point estimate $\hat{X}_{SM}$, denoted $\hat{\sigma}$($\hat{X}_{SM}$), equals $\hat{\sigma}(\hat{X}_{SM}) = \sqrt{ \sum_{i} N_i \left(N_i - n_i\right) \frac{s_{x_i}^2}{n_i} }$. The achieved precision ($A'_{SM}$) of the point estimate $\hat{X}_{SM}$ equals $t_{0.05,\text{DF}}\hat{\sigma}(\hat{X}_{SM})$ where degrees of freedom (DF) are given by the Satterthwaite approximation formula: $\text{DF} = \frac{\left( \sum_i N_i (N_i - n_i) \frac{s_{x_i}^2}{n_i} \right)^2}
{\sum_i      \left[ N_i (N_i - n_i) \frac{s_{x_i}^2}{n_i} \right]^2/(n_i - 1)}$. The stratified mean one-sided 95\% LCB on total QREs in the sampling frame equals $\text{LCB}_{SM}(\hat{T_Q})$ = $\hat{X}_{SM}$ - $A'_{SM}$.\\\\
\textbf{\underline{Stratified Difference Estimator}}\\\\
The average difference ($\bar{d}_i$) between actual QREs and pQREs within the $i^{\text{th}}$ stratum, is estimated by $\bar{d}_i = \frac{\sum_{j=1}^{n_i} d_{ij}}{n_i}$. In this notation,  $d_{ij}$ is the difference between actual QREs ($x_{ij}$) and pQREs ($y_{ij}$) for the $j^{\text{th}}$ sampled project in the $i^{\text{th}}$ stratum.  A point estimate of the total difference between actual QREs and pQREs is $\hat{D}_{S} = \sum_{i} N_{i} \bar{d}_i$. A point estimate of $T_Q$, actual QREs for the sampling frame, is $\hat{T_Q} = \hat{X}_{SD}$ = T + $\hat{D}_{S}$, where T equals total pQREs. \\\\
The sample variance ($s_{d_i}^2$) among differences within each stratum equals $s_{d_i}^2 = \frac{ \sum_{j=1}^{n_i} d_{ij}^2 - n_i \bar{d}_i^2}{n_i - 1} $. The standard error of the point estimate $\hat{X}_{SD}$ equals $\hat{\sigma}(\hat{D}_{S}) = \sqrt{ \sum_{i} N_i \left(N_i - n_i\right) \frac{s_{d_i}^2}{n_i} }$ since total pQREs (T) is a constant. The achieved precision ($A'_{SD}$) of the point estimate $\hat{X}_{SD}$ equals $t_{0.05,\text{DF}}\hat{\sigma}(\hat{D}_{S})$ where degrees of freedom (DF) are given by the Satterthwaite approximation formula: $\text{DF} = \frac{\left( \sum_i N_i (N_i - n_i) \frac{s_{d_i}^2}{n_i} \right)^2}
{\sum_i      \left[ N_i (N_i - n_i) \frac{s_{d_i}^2}{n_i} \right]^2/(n_i - 1)}$. The stratified difference one-sided 95\% LCB on total QREs in the sampling frame equals $\text{LCB}_{SD}(\hat{T_Q})$ = $\hat{X}_{SD}$ - $A'_{SD}$.
\subsection*{\text {Simulation Study}}
We conduct a simulation study to evaluate the performance of the five different estimators used to produce a one-sided 95\% LCB on total QREs for a sampling frame.
\subsubsection*{\text {Sampling Frame}}
The sampling frame for the simulation consists of 921 potentially qualified R\&D projects, having pQREs totaling \$20,932,340.21.  The pQREs in the sampling frame follow a typical right-skewed distribution as depicted in the histogram in Figure 1.\\
\begin{figure}[h!]
\centering
\includegraphics[width=1\linewidth]{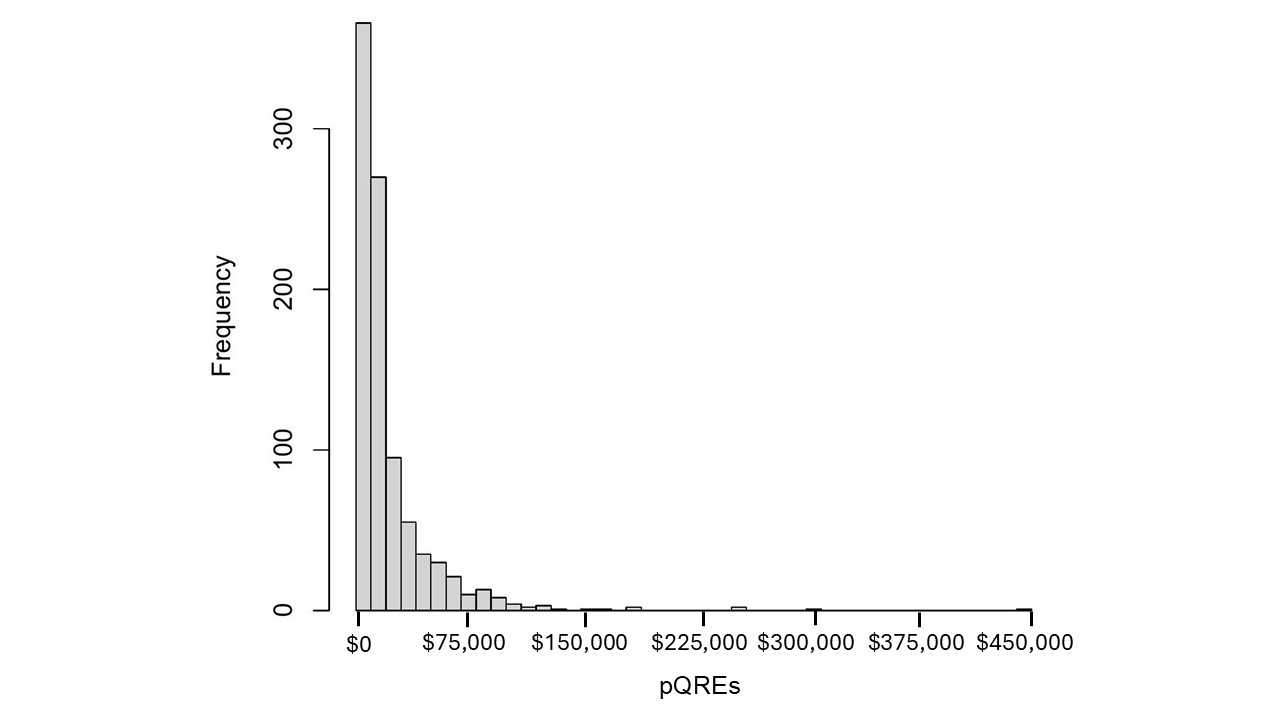}
\caption{Histogram of pQREs in the Sampling Frame (N = 921)}
\label{fig:pqre_hist}
\end{figure}

\noindent The sampling frame is segmented into k = 4 strata as depicted in Table 2.
\begin{table}[h!]
\centering
\caption{Stratified Population of Potentially Qualified R\&D Projects (N = 921)}
\label{tab:strata}
\begin{tabular}{|c|c|c|c|c|}
\hline
\textbf{Strata} & \textbf{N} & \textbf{n} & \textbf{Total pQREs} & \textbf{Average pQREs/Project} \\
\hline
1 & 4   & 4     & \$1{,}256{,}601.37 & \$314{,}150.34 \\
\hline
2 & 51  & $n_2$ & \$4{,}852{,}064.46 & \$95{,}138.52 \\
\hline
3 & 180 & $n_3$ & \$7{,}283{,}339.10 & \$40{,}463.00 \\
\hline
4 & 686 & $n_4$ & \$7{,}540{,}335.28 & \$10{,}991.74 \\
\hline
\textbf{TOTAL} & \textbf{921} & \textbf{$\sum_i n_i$} & \textbf{\$20{,}932{,}340.21} & \textbf{---} \\
\hline
\end{tabular}
\end{table}
\par \noindent Given the large average pQREs of \$314,150.34, among the four projects in Stratum 1, these projects are assigned to a certainty strata and assumed to be 100\% qualified.  Accordingly, the simulation study reflects variation in the LCB estimates on total QREs arising solely from the remaining 917 projects in Strata 2–4. Henceforth, the term sampling frame refers to projects across Strata 2-4.\\\\
Although pQREs are known for every project in the sampling frame, the qualification status of each project is unknown.  The simulation considers a range of scenarios in which the proportion of qualified projects in Strata 2–4 is specified by the parameter vector ($p_2$, $p_3$, $p_4$).  Table 3 presents the range of assumed values of ($p_2$, $p_3$, $p_4$), the resulting average value of $T_Q$ and the average parameters $\mu_{T}$, $\mu_{Q}$ and $\mu_{NQ}$ across Strata 2-4. Since the certainty stratum, Stratum 1, does not influence variability of the LCB estimators, the average parameters $\mu_{T}$, $\mu_{Q}$ and $\mu_{NQ}$ are computed across Strata 2-4 only.\\
\begin{table}[h!]
\centering
\caption{Assumed Parameter Vector ($p_2$, $p_3$, $p_4$) and E[$T_Q$], $\mu_{T}$, $\mu_{Q}$ and $\mu_{NQ}$.}
\label{tab:strata}
\begin{tabular}{|c|c|c|c|c|}
\hline
\textbf{($p_2$, $p_3$, $p_4$)} & \textbf{E[$T_Q$]} & \textbf{$\mu_{T,2-4}$} & \textbf{$\mu_{Q,2-4}$} & \textbf{$\mu_{NQ,2-4}$}  \\
\hline
\textbf{Scenario 1} & & & & \\ 
\hline
\{0.95,0.95,0.95\} & \$19{,}948{,}553 & \$21{,}457 &  \$21{,}457 & \$21{,}457 \\
\hline
\{0.85,0.85,0.85\} & \$17{,}980{,}979 & \$21{,}457 &  \$21{,}457 & \$21{,}457 \\
\hline
\{0.75,0.75,0.75\} & \$16{,}013{,}405 & \$21{,}457 &  \$21{,}457 & \$21{,}457 \\
\hline
\{0.65,0.65,0.65\} & \$14{,}045{,}832 & \$21{,}457 &  \$21{,}457 & \$21{,}457 \\
\hline
\{0.5,0.5,0.5\} & \$11{,}094{,}471 & \$21{,}457 &  \$21{,}457 & \$21{,}457 \\
\hline
\{0.25,0.25,0.25\} & \$6{,}175{,}536 & \$21{,}457 &  \$21{,}457 & \$21{,}457 \\
\hline
\textbf{Scenario 2} & & & & \\ 
\hline
\{0.95,0.8,0.7\} & \$16{,}970{,}969 & \$21{,}457 & \$23{,}362 &  \$16{,}212\\
\hline
\{0.95,0.7,0.5\} & \$14{,}734{,}568 & \$21{,}457 & \$26{,}047 &  \$15{,}512\\
\hline
\{0.85,0.6,0.4\} & \$12{,}766{,}994 & \$21{,}457 & \$27{,}036 &  \$16{,}622\\
\hline
\{0.75,0.6,0.4\} & \$12{,}281{,}787 & \$21{,}457 & \$26{,}210 &  \$17{,}428\\
\hline
\{0.75,0.6,0.25\} & \$11{,}150{,}737 & \$21{,}457 & \$31{,}138 &  \$16{,}323\\
\hline
\{0.85,0.5,0.25\} & \$10{,}907{,}610 & \$21{,}457 & \$31{,}658 & \$16{,}376 \\
\hline
\textbf{Scenario 3} & & & & \\ 
\hline
\{0.7,0.8,0.95\} & \$17{,}643{,}036 & \$21{,}457 & \$19{,}709 &  \$38{,}426\\
\hline
\{0.5,0.7,0.95\} & \$15{,}944{,}289 & \$21{,}457 & \$18{,}286 &  \$43{,}832\\
\hline
\{0.4,0.6,0.85\} & \$13{,}976{,}716 & \$21{,}457 & \$17{,}878 &  \$33{,}847\\
\hline
\{0.4,0.6,0.75\} & \$13{,}222{,}682 &  \$21{,}457 & \$18{,}613 & \$28{,}127 \\
\hline
\{0.25,0.5,0.85\} & \$12{,}520{,}572 & \$21{,}457 & \$16{,}423 &  \$36{,}391\\
\hline
\{0.25,0.6,0.75\} & \$12{,}494{,}872 & \$21{,}457 & \$17{,}691 & \$29{,}947 \\
\hline
\end{tabular}
\end{table}
\par \noindent The R = 10,000 simulation runs proceed as follows:
\begin{enumerate}
\item Assign a qualification status to each project in the sampling frame according to the specified values of ($p_2$, $p_3$, $p_4$).
\item Compute $T_Q$, actual QREs, by zeroing out pQREs for all non-qualified projects.
\item Draw a random sample of \emph{n} projects.  For a stratified random sample, \emph{n} = $\sum_{i=1}^{k} n_i$ and each stratum is sampled independently.
\item Estimate the one-sided 95\% LCB, $\text{LCB}_{\mathbf{\cdot}}(\hat{T_Q})$ for the selected estimator (i.e. SA, SSA, CSA, SM or SD).
\item Compute the ratio $\frac{\text{LCB}_{\mathbf{\cdot}}(\hat{T_Q})}{T_Q}$.
\item Record an indicator equal to one when the LCB fails, i.e. when $\text{LCB}_{\mathbf{\cdot}}(\hat{T_Q}) > T_Q$, and zero otherwise.
\end{enumerate}
Across all R = 10,000 simulation runs, the average ratio of $\frac{\text{LCB}_{\mathbf{\cdot}}(\hat{T_Q})}{T_Q}$ and the percentage of times the LCB fails is reported.  A simulation is conducted for each estimator separately and each assigned parameter vector ($p_2$, $p_3$, $p_4$). The performance of $\text{LCB}_{SM}(\hat{T_Q})$ and $\text{LCB}_{SD}(\hat{T_Q})$ are also assessed in a combined scenario in which both LCBs are computed in parallel and the LCB associated with the smaller standard error is selected (consistent with IRS Revenue Procedure 2011-42). In this combined scenario, the percentage of runs in which $\text{LCB}_{SD}(\hat{T_Q})$ is chosen is also recorded. Two sample sizes are considered:  n = 30 ($n_1$ = 4, $n_2$ = 9, $n_3$ = 9, $n_4$ = 8) and n = 18 ($n_1$ = 4, $n_2$ = 5, $n_3$ = 5, $n_4$ = 4).\\\\
Because the convolution-based stratified attribute estimator $\text{LCB}_{CSA}(\hat{T_Q})$ utilizes a computationally intensive branch-and-bound search algorithm over the $\mathcal{A}_{0.95}$ parameter space, a lookup table listing $\text{LCB}_{CSA}(\hat{T_Q})$ for each possible value of \emph{e} is constructed and incorporated into the simulation. \\\\
In a separate simulation, we evaluate the effect of resampling on the stratified mean and difference estimators.  Step 3 of the simulation is modified as follows:
\begin{itemize}
\item[] 3A) Draw 50 stratified random samples of \emph{n} projects, where \emph{n} = $\sum_{i=1}^{k} n_i$ and each stratum is sampled independently.
\item[] 3B) Compute $\hat{T} = \sum_{i=1}^{k} N_i \times \bar{y}_i$, where $\bar{y}_i = \frac{\sum_{j=1}^{n_i} y_{ij}}{n_i}$, the average pQREs among sampled projects in the $i^{\text{th}}$ stratum. 
\item[] 3C) Retain the random sample with the largest $\hat{T}$ and discard the remaining 49 samples.
\end{itemize}
As before, two sample sizes are considered: n = 30 ($n_1$ = 4, $n_2$ = 9, $n_3$ = 9, $n_4$ = 8) and n = 18 ($n_1$ = 4, $n_2$ = 5, $n_3$ = 5, $n_4$ = 4).  
\section*{\text {Results}}
Tables 4-9 report, for each of the five LCB estimators and each sampling frame structure, the percentage of simulation runs (R = 10,000) in which the estimated LCB fails (i.e. $\text{LCB}_{\mathbf{\cdot}}(\hat{T_Q}) > T_Q$) and the average ratio $\frac{\text{LCB}_{\mathbf{\cdot}}(\hat{T_Q})}{T_Q}$.  Results are reported separately for sample sizes \emph{n} = 30 and \emph{n} = 18. When the LCB fails in more than 5\% of the simulation runs, the estimator does not produce a valid one-sided 95\% LCB on actual QREs for that sampling frame structure. The average ratio $\frac{\text{LCB}_{\mathbf{\cdot}}(\hat{T_Q})}{T_Q}$ subtracted from one reflects the average "haircut" on actual QREs (not pQREs) for that LCB estimator.  For reference, an unbiased point estimator of $T_Q$ ($\hat{T_Q}$) would yield an average ratio $\frac{\hat{T_Q}}{T_Q}$ near one and an average "haircut" on actual QREs near zero.\\\\
Table 4 (\emph{n} = 30) and Table 5 (\emph{n} = 18) depict the performance of the simple attribute (SA), simple stratified attribute (SSA) and convolution-based stratified attribute (CSA) LCB estimators for each sampling frame structure. For n = 30, the $\text{LCB}_{SA}$ estimator produced a valid and conservative one-sided 95\% LCB for every sampling frame structure under Scenarios 1 and 2, where failure frequencies of $\text{LCB}_{SA}$ ranged from 0\% to 3.6\%. However, failure frequencies of $\text{LCB}_{SA}$ ranged from 16.9\% to as high as 51.9\% under Scenario 3. For the qualification parameter vector ($p_2$ = 0.25, $p_3$ = 0.5, $p_4$ = 0.85) under Scenario 3, the average ratio $\frac{\text{LCB}_{SA}(\hat{T_Q})}{T_Q}$ was 1.007, the average ratio expected for a point estimator of $T_Q$, not a LCB.\\\\
For n = 30, the failure frequencies of the $\text{LCB}_{SSA}$ estimator decreased relative to the $\text{LCB}_{SA}$ estimator under Scenario 3, ranging from 0.3\% to 1.0\%.
However, the failure frequencies of the $\text{LCB}_{SSA}$ estimator increased relative to the $\text{LCB}_{SA}$ estimator under Scenario 2, ranging from 3.8\% to 6.7\%.
The $\text{LCB}_{CSA}$ estimator produced failure frequencies under 5\% for every sampling frame structure, ranging from 0\% to 3.2\% across Scenarios 1-3. Table 5 (n = 18) depicted similar trends although average ratios $\frac{\text{LCB}_{\mathbf{\cdot}}(\hat{T_Q})}{T_Q}$ were smaller (i.e. larger "haircut" on actual QREs). For n = 18, failure frequencies of $\text{LCB}_{SA}$ reduced, ranging from 2.7\% to 28.6\% under Scenario 3.\\\\
Table 6 (n = 30) and Table 7 (n = 18) depict the performance of the stratified mean (SM) and stratified difference (SD) LCB estimators for each sampling frame structure without resampling. For n = 30, the $\text{LCB}_{SM}$ estimator produced failure frequencies under 5\% for every sampling frame structure, ranging from 1.9\% to 4.9\% across Scenarios 1-3. In contrast, the $\text{LCB}_{SD}$ estimator produced failure frequencies above 5\% for every sampling frame structure, ranging from 5.7\% to 28.3\% across Scenarios 1-3. The highest failure frequencies of the $\text{LCB}_{SD}$ estimator occurred in sampling frame structures under Scenario 1, in particular where qualification probabilities were high. For qualification parameter vectors ($p_2$ = 0.95, $p_3$ = 0.95, $p_4$ = 0.95) and ($p_2$ = 0.85, $p_3$ = 0.85, $p_4$ = 0.85), failure frequencies of the $\text{LCB}_{SD}$ estimator were 28.3\% and 14.1\%, respectively.\\\\
Table 6 also depicts the performance of the combined scenario, whereby $\text{LCB}_{SM}$ and $\text{LCB}_{SD}$ are computed in parallel and the LCB associated with the smaller standard error is selected. For \emph{n} = 30, the combined scenario also produced failure frequencies above 5\% for all but one sampling frame structure, ranging from 4.9\% to 27.7\% across Scenarios 1-3, in line with the failure frequencies of the $\text{LCB}_{SD}$ estimator. Table 7 (n = 18) depicts similar trends although the average ratios $\frac{\text{LCB}_{\mathbf{\cdot}}(\hat{T_Q})}{T_Q}$ were smaller (i.e larger "haircut" on actual QREs). However, $\text{LCB}_{SD}$ failure frequencies increased, particularly under Scenario 1 where qualification probabilities were high.\\\\
Table 8 (n = 30) and Table 9 (n = 18) depict the performance of the stratified mean (SM) and stratified difference (SD) LCB estimators for each underlying sampling frame structure with resampling. For n = 30, the $\text{LCB}_{SM}$ estimator produced failure frequencies above 5\% for all but one sampling frame structure, ranging from 4.5\% to 50.4\% across Scenarios 1-3. The highest failure frequency for the $\text{LCB}_{SM}$ estimator was 50.4\% and occurred under Scenario 1 for the qualification parameter vector ($p_2$ = 0.95, $p_3$ = 0.95, $p_4$ = 0.95). The average "haircut" on actual QREs, reflected by one minus the average ratio $\frac{\text{LCB}_{SM}(\hat{T_Q})}{T_Q}$ similarly decreased under resampling.\\\\
Under resampling (n = 30), the $\text{LCB}_{SD}$ estimator produced failure frequencies lower than the corresponding failure frequencies in Table 6, ranging from 0.3\% to 27.6\% across Scenarios 1-3. The average "haircut" on actual QREs increased under resampling compared to Table 6. The combined scenario produced failure frequencies similarly aligned with the failure frequencies of the $\text{LCB}_{SD}$ estimator, ranging from 1.9\% to 27.6\%. Table 9 (n = 18) depicted similar trends under resampling although the average ratios $\frac{\text{LCB}_{\mathbf{\cdot}}(\hat{T_Q})}{T_Q}$ were smaller (i.e larger "haircut" on actual QREs) compared to Table 8. For n = 18, failure frequencies of $\text{LCB}_{SD}$ increased relative to Table 8, ranging from 0.5\% to 49.4\% across Scenarios 1-3. 
\section*{Discussion}
The domestic R\&D tax credit provides one of the largest economic tax incentives in the United States, generating billions of dollars in corporate tax savings each year. To claim the credit, companies must carefully document and accurately determine total QREs, often a complex undertaking involving diverse business components. Statistical sampling offers a powerful way to simplify QRE estimation and can enable more companies to efficiently obtain the credit.\\\\
Our study is the first to evaluate, side-by-side, the performance of the prevailing statistical methods outlined in IRS Revenue Procedure 2011-42, namely the simple attribute estimator and both the stratified mean and difference variable estimators. As outlined in Table 1, these three statistical methods approach the common goal of QRE estimation from distinct theoretical frameworks and typically produce different total QRE estimates for the same random sample. Very large high-pQRE projects should always be sampled directly in order to minimize sampling variability.  This recommended practice was incorporated into our simulations, irrespective of the statistical method.\\\\
The simple attribute estimator ($\text{LCB}_{SA}$) computes a one-sided 95\% LCB (r) on the proportion of qualified projects in the sampling frame and rescales pQREs (T) according to $\text{LCB}_{SA} = rT$.  The one-sided 95\% LCB (r) is derived from the exact hypergeometric probability distribution and is valid for any sample size. However, the validity of $\text{LCB}_{SA}$ as a one-sided 95\% LCB on total QREs depends on how non-qualified projects are distributed within the sampling frame. The $\text{LCB}_{SA}$ estimator produced a valid and conservative one-sided 95\% LCB on total QREs across all Scenario 1 and 2 sampling frame structures. However, under Scenario 3, failure frequencies "blew up" to over 50\% and the average "haircut" on actual QREs approached zero. \\\\
Each of the Scenario 3 sampling frame structures outlined in Table 3 indicate that $\mu_{Q} < \mu_{T} < \mu_{NQ}$ because the small number of high pQRE projects in Stratum 2 were more likely to be non-qualified than lower pQRE projects in Strata 3 and 4. A simple random sample will identify fewer projects from Stratum 2 relative to the more populous Strata 3 and 4. Furthermore, the "haircut" on pQREs for a $\text{LCB}_{SA}$ estimator is $(1-r)T$, algebraically equivalent to z*$\times \mu_{T}$. Because ${\mu}_T$ represents average pQREs across all projects in the sampling frame, this "haircut" does not account for the disproportionately higher pQREs associated with non-qualified Stratum 2 projects.\\\\
A stratified random sampling design ensures adequate coverage of high pQRE projects in Stratum 2. As a result, failure frequencies for the simple stratified attribute $\text{LCB}_{SSA}$ estimator were below 5\% for each sampling frame structure under Scenario 3. Failure frequencies increased modestly above 5\% in sampling frame structures under Scenario 2 relative to the $\text{LCB}_{SA}$ estimator. Under Scenario 2, lower pQRE non-qualified projects in Strata 4 are proportionally under-represented in the stratified random sample.\\\\
These simulation results for the $\text{LCB}_{SSA}$ estimator underscore the advantages of a stratified random sample, but should be interpreted with care. The SSA method is “ad hoc”, in that the rescaling factor r does not represent a valid one-sided 95\% LCB on the proportion of qualified projects in the sampling frame as it did for the SA method. Notably, sampling high pQRE projects also offers a practical way to rule out Scenario 3 sampling frame structures early on in a study or serve as a reflex sample to confirm that the $\text{LCB}_{SA}$ estimator resulted in a valid one-sided 95\% LCB on actual QREs.\\\\
The convolution-based stratified attribute $\text{LCB}_{CSA}$ estimator produced failure frequencies below 5\% for all sampling frame structures across Scenarios 1-3. The $\text{LCB}_{CSA}$ estimator is unique in that it recognizes a parameter vector $\theta$ which includes $N_{NQ,i}$ for each of the \emph{i} strata, thereby accounting for the possibility of varying qualification probabilities among strata. Over the 95\% acceptance region of $\theta$, $\mathcal{A}_{0.95}$, the branch-and-bound search algorithm identifies the minimum $T_Q (\boldsymbol{\theta})$, thereby directly computing a one-sided 95\% LCB on total QREs in the sampling frame.\\\\
The novel $\text{LCB}_{CSA}$ estimator offers a practical addition to the QRE statistical estimation toolkit and has several advantages over variable methods. As an attribute method, the $\text{LCB}_{CSA}$ estimator relies on detected erroneous, non-qualified projects in a stratified random sample. These are unknown at the outset of an R\&D study and therefore the $\text{LCB}_{CSA}$ estimator is not susceptible to resampling bias. Determination of $\mathcal{A}_{0.95}$ relies on the exact hypergeometric probability distribution, and produces a valid one-sided 95\% LCB on actual QREs, even for small sample sizes. The branch-and-bound search algorithm used to identify the minimum $T_Q (\boldsymbol{\theta})$ over $\mathcal{A}_{0.95}$ is efficient and feasible for routine application in R\&D studies. \\\\
A caveat to our simulation study is that we modeled "between-strata" variability in qualification probabilities, but not "within-strata" variability.  Were a tendency for Scenario 3 behavior to arise within a given strata, the $\text{LCB}_{CSA}$ estimator could yield failure frequencies higher than 5\%. As an attribute method, $\text{LCB}_{CSA}$ estimator also does not allow for “shrink-back” of pQREs, whereby pQREs for a sampled project may be partially allotted. \\\\ 
Variable methods for QRE estimation are viewed favorably by the IRS due to the direct nexus between sampled projects and QREs. In our study, the stratified mean estimator $\text{LCB}_{SM}$ did perform well independently and produced failure frequencies under 5\% for all sampling frame structures across Scenarios 1-3. However, the $\text{LCB}_{SM}$ estimator was also the most susceptible to overestimation of QREs due to resampling bias.  Under resampling, failure frequencies for the $\text{LCB}_{SM}$ estimator rose above 50\%, with the highest failure frequencies occurring in sampling frame structures having the highest qualification probabilities.\\\\
The $\text{LCB}_{SM}$ estimator is particularly susceptible to resampling bias because actual QREs are projected directly from the sampled projects.  Unlike with "haircut"-based approaches, total pQREs (T) do not serve as a hard upper bound on a final QRE estimate.  Samples with higher than average pQREs are easily pre-selected and other random samples discarded. No safeguards exist in the IRS guidance against resampling, only a requirement to report the random seed for the selected sample.  Discarded samples are undetectable. Resampling bias is mitigated in sampling frame structures with low qualification probabilities since qualification status is unknown at the outset of an R\&D study.\\\\
The stratified difference $\text{LCB}_{SD}$ estimator produced failure frequencies over 5\% for all sampling frame structures across Scenarios 1-3.  The distribution of differences between actual QREs and pQREs is right-skewed, and particularly so when qualification probabilities are high and there is a large preponderance of zero differences in the sampling frame. For highly right-skewed distributions, a sample size of n = 30 is inadequate to meet the asymptotic conditions of the Central Limit Theorem and guarantee the sampling distribution of the point estimate of the total difference $\hat{D}_{S}$ is normally distributed. Construction of a valid one-sided 95\% LCB on total QREs relies on the assumption that $\hat{D}_{S}$ is normally distributed.\\\\
The $\text{LCB}_{SD}$ estimator can overestimate actual QREs when fewer than expected erroneous, non-qualified projects are identified in a sample, reducing both the magnitude of $\hat{D}_{S}$ and its estimated standard error, $\hat{\sigma}(\hat{D}_{S})$ and thereby underestimating the resulting "haircut" on pQREs. In the extreme case, zero erroneous, non-qualified projects are sampled and the $\text{LCB}_{SD}$ estimator produces total pQREs (T). In our simulations, the $\text{LCB}_{SD}$ estimator produced lower failure frequencies when qualification probabilities were lower. Fewer qualified projects in the sampling frame reduces right-skewness in the sampling distribution of $\hat{D}_{S}$ and lowers the chance that zero erroneous, non-qualified projects are sampled. Nonetheless, the $\text{LCB}_{SD}$ estimator is a "haircut"-based approach and cannot exceed total pQREs (T). As a result, over-estimation of pQREs by the $\text{LCB}_{SD}$ estimator, when it does occur, is bounded. \\\\
Under resampling, the $\text{LCB}_{SD}$ estimator failure probabilities decreased.  Selection of higher pQRE samples will inflate differences among the sampled non-qualified projects, thereby increasing the estimated standard error $\hat{\sigma}(\hat{D}_{S})$ of the $\text{LCB}_{SD}$ estimator. Larger values of $\hat{\sigma}(\hat{D}_{S})$ increase the average "haircut" on pQREs, decreasing the likelihood of overestimating total QREs.\\\\
We implemented a combined scenario in which the $\text{LCB}_{SM}$ and $\text{LCB}_{SD}$ estimators are computed in parallel and the estimator with the smaller standard error is selected.  A perhaps unintended consequence of this "standard error rule" is that when fewer than expected erroneous, non-qualified projects are identified in a sample, the $\text{LCB}_{SD}$ is more likely to over-estimate $T_Q$ and have the smaller estimated standard error $\hat{\sigma}(\hat{D}_{S})$ and be chosen. Because $\text{LCB}_{SD}$ failures coincide with its selection, failure frequencies for the combined scenario were closely aligned with independent failure frequencies of $\text{LCB}_{SD}$, with or without resampling.\\\\
Interestingly, the "standard error rule" also mitigated the strongest effects of resampling bias on the $\text{LCB}_{SM}$ estimator which was highest when qualification probabilities were high. When qualification probabilities are high, the $\text{LCB}_{SD}$ estimator was more frequently selected.  We opted not to implement the "relative precision rule" in our simulations which allows the point estimator to replace the LCB when the relative precision is less than 10\%.  Because this rule is “ad hoc”, it does not meaningfully reveal the statistical properties of the various estimators under consideration.\\\\
In conclusion, the prevailing statistical methods used to estimate a one-sided 95\% LCB on total QREs outlined in IRS Revenue Procedure 2011-42 are imperfect and poorly understood. Each of these methods produced inflated failure frequencies above 5\% in some or all of our simulations and failure frequencies varied by sampling frame structure. In cases of a contested credit, the Tax Court has relied on mutually agreed-upon samples rather than random samples that can produce statistically rigorous results. Our study aims to strengthen the utility of statistical methods in QRE estimation by clarifying their relative strengths and weaknesses. Furthermore, our novel convolution-based stratified attribute method provides a fresh and practical addition to the QRE statistical estimation toolbox.  
\section*{References}
\begin{enumerate}
\item One Big Beautiful Bill Act, H.R. 1, 119th Cong. (2025).
\item \emph{George v. Commissioner}, T.C. Memo 2026-10.
\item \emph{Smith v. Commissioner}, T.C. Memo 2026-50.
\item Song T., Ashtiani H., Seo J., St. Martin D., Benton K., Rotz W. (2022). "The research credit: Using statistical sampling." \emph{The Tax Adviser}, 53(2): 11-16. https://www.thetaxadviser.com/issues/2022/feb/research-credit-using-statistical-sampling/.
\item Internal Revenue Code, 26 U.S.C. \S 41(d)(1) (2024).
\item Internal Revenue Service. (2026). \emph{Internal Revenue Manual} (IRM \S 4.47.3).
\item Rev. Proc. 2011-42, 2011-42 I.R.B. 518.
\item Ashtiani H., Benton K., Rotz W. (2025). "Tax court declines to limit discovery for sampled research credit claims." \emph{The Tax Adviser}, 56(2).\\  https://www.thetaxadviser.com/issues/2025/feb/tax-court-limits-discovery-for-sampled-research-credit-claims/.
\item \emph{Suder v. Commissioner}, T.C. Memo 2014-201.
\item \emph{Kapur v. Commissioner}, T.C. Memo 2024-28.
\item \emph{Bayer Corp. \& Subsidiaries v. United States}.  Civil Action No. 09-351 (W.D. Pa. 2012). https://www.govinfo.gov/app/details/USCOURTS-pawd-2\_09-cv-00351.
\item Cochran, W. G. (1977). Sampling techniques (3rd ed.). John Wiley \& Sons.
\item Internal Revenue Service. 2026. \emph{Form 6765: Credit for Increasing Research Activities}. U.S. Department of the Treasury. www.irs.gov.
\item \emph{Leon Max v. Commissioner}, T.C. Memo 2021-37.
\item \emph{Felker v. Commissioner}, Docket No. 3871-17 (T.C. settled).
\item \emph{Intermountain Electronics Inc. v. Commissioner},  Docket No. 11019-19 (T.C.).
\end{enumerate}
\newpage
\begin{table}[ht]
\centering
{\scriptsize
\setlength{\tabcolsep}{4pt}
\renewcommand{\arraystretch}{1}
\section*{Tables}
\begin{tabular}{|l|c|c|c|c|c|c|c|}
\hline  
& & \multicolumn{2}{c}{\textbf{Non-Stratified}} & \multicolumn{4}{c}{\textbf{Stratified Results}} \\
\cline{5-8}
 &
 &
\multicolumn{2}{c}{\textbf{Simple Attribute}} &
\multicolumn{2}{c}{\textbf{Simple Attribute}} &
\multicolumn{2}{c}{\textbf{Convolution}}
\\
\cline{3-4} \cline{5-6} \cline{7-8} 
\textbf{\{$p_2,p_3,p_4$\}} & \textbf{E[$T_Q$]} & \% LCB Fails & E[$\frac{\text{LCB}_{SA}(\hat{T_Q})}{T_Q}$] & \% LCB Fails & E[$\frac{\text{LCB}_{SSA}(\hat{T_Q})}{T_Q}$] & \% LCB Fails & E[$\frac{\text{LCB}_{CSA}(\hat{T_Q})}{T_Q}$] \\
\hline
\textbf{Scenario 1} & & & & & & & \\
\{0.95,0.95,0.95\}  & \$19,948,553  & 0\% & 0.8708 & 0\% & 0.8703 & 0\% & 0.8491  \\
\{0.85,0.85,0.85\} & \$17,980,979 & 2.0\% & 0.8289 & 1.7\% & 0.8258 & 1.2\% & 0.7931  \\
\{0.75,0.75,0.75\} & \$16,013,405 & 3.6\% & 0.7910 & 3.1\% & 0.7911 & 1.3\% & 0.7498  \\
\{0.65,0.65,0.65\} & \$14,045,832 & 3.3\% & 0.7548 & 3.0\% & 0.7548 & 1.2\% & 0.7049 \\
\{0.5,0.5,0.5\} & \$11,094,471 & 3.4\% & 0.7021 & 3.1\% & 0.7053 & 1.2\% & 0.6470  \\
\{0.25,0.25,0.25\} & \$6,175,536 & 3.4\% & 0.6032 & 3.0\% & 0.6031 & 1.3\% & 0.5447  \\
\textbf{Scenario 2} & & & & & & & \\
\{0.95,0.8,0.7\} & \$16,970,969 & 0.5\% & 0.7269 & 3.8\% & 0.8364 & 2.5\% & 0.8001  \\
\{0.95,0.7,0.5\} & \$14,734,568 & 0.1\% & 0.6115 & 5.8\% & 0.8222 & 2.9\% & 0.7782  \\
\{0.85,0.6,0.4\} & \$12,766,994 & 0.1\% & 0.5608 & 5.6\% & 0.7902 & 2.7\% & 0.7411  \\
\{0.75,0.6,0.4\} & \$12,281,787 & 0.2\% & 0.5750 & 5.0\% & 0.7688 & 2.2\% & 0.7148 \\
\{0.75,0.6,0.25\} & \$11,150,737 & 0.1\% & 0.4653 & 5.5\% & 0.7686 & 2.6\% & 0.7144  \\
\{0.85,0.5,0.25\} & \$10,907,610 & 0.02\% & 0.4551 & 6.7\% & 0.7862 & 3.2\% & 0.7296  \\
\textbf{Scenario 3} & & & & & & &  \\
\{0.7,0.8,0.95\} & \$17,643,036 & 16.9\% & 0.9210 & 0.7\% & 0.7915 & 0.4\% & 0.7588  \\
\{0.5,0.7,0.95\} & \$15,944,289 & 39.5\% & 0.9715 & 0.6\% & 0.7376 & 0.2\% & 0.6937 \\
\{0.4,0.6,0.85\} & \$13,976,716 & 34.1\% & 0.9462 & 0.7\% & 0.7007 & 0.2\% & 0.6520  \\
\{0.4,0.6,0.75\} & \$13,222,682 & 19.6\% & 0.8813 & 1.0\% & 0.6977 & 0.3\% & 0.6470  \\
\{0.25,0.5,0.85\} & \$12,520,572 & 51.9\% & 1.007 & 0.3\% & 0.6504 & 0.1\% & 0.6028  \\
\{0.25,0.6,0.75\} & \$12,494,872 & 28.7\% & 0.9184 & 0.4\% & 0.6588 & 0.1\% & 0.6091  \\
\hline
\end{tabular}
}
\caption{Simulation results (R = 10,000) for attribute estimators (SA, SSA and CSA) ($n_1 = 4, n_2 = 9, n_3 = 9, n_4 = 8$).}
\label{tab:sim_results_grouped}
\end{table}
\newpage
\begin{table}[ht]
\centering
{\scriptsize
\setlength{\tabcolsep}{4pt}
\renewcommand{\arraystretch}{1}

\begin{tabular}{|l|c|c|c|c|c|c|c|}
\hline  
& & \multicolumn{2}{c}{\textbf{Non-Stratified}} & \multicolumn{4}{c}{\textbf{Stratified Results}} \\
\cline{5-8}
 &
 &
\multicolumn{2}{c}{\textbf{Simple Attribute}} &
\multicolumn{2}{c}{\textbf{Simple Attribute}} &
\multicolumn{2}{c}{\textbf{Convolution}}
\\
\cline{3-4} \cline{5-6} \cline{7-8} 
\textbf{\{$p_2,p_3,p_4$\}} & \textbf{E[$T_Q$]} & \% LCB Fails & E[$\frac{\text{LCB}_{SA}(\hat{T_Q})}{T_Q}$] & \% LCB Fails & E[$\frac{\text{LCB}_{SSA}(\hat{T_Q})}{T_Q}$] & \% LCB Fails & E[$\frac{\text{LCB}_{CSA}(\hat{T_Q})}{T_Q}$] \\
\hline
\textbf{Scenario 1} & & & & & & & \\
\{0.95,0.95,0.95\} & \$19,948,553 & 0\% & 0.7933 & 0\% & 0.7921 & 0\% & 0.7656  \\
\{0.85,0.85,0.85\} & \$17,980,979 & 0.1\% & 0.7425 & 0.1\% & 0.7425 & 0.04\% & 0.7082  \\
\{0.75,0.75,0.75\} & \$16,013,405 & 1.9\% & 0.7016 & 1.8\% & 0.6977 & 1.6\% & 0.6595  \\
\{0.65,0.65,0.65\} & \$14,045,832 & 2.4\% & 0.6601 & 2.3\% & 0.6579 & 1.6\% & 0.6136 \\
\{0.5,0.5,0.5\} & \$11,094,471 & 2.8\% & 0.5971 & 2.6\% & 0.5988 & 1.6\% & 0.5481  \\
\{0.25,0.25,0.25\} & \$6,175,536 & 3.3\% & 0.4993 & 2.8\% & 0.4953 & 1.4\% & 0.4556  \\
\textbf{Scenario 2} & & & & & & & \\
\{0.95,0.8,0.7\} & \$16,970,969 & 0.9\% & 0.6395 & 4.1\% & 0.7547 & 0.8\% & 0.7165  \\
\{0.95,0.7,0.5\} & \$14,734,568 & 0.3\% & 0.5227 & 5.0\% & 0.7315 & 1.6\% & 0.6897  \\
\{0.85,0.6,0.4\} & \$12,766,994 & 0.2\% & 0.4733 & 4.7\% & 0.6962 & 2.0\% & 0.6436  \\
\{0.75,0.6,0.4\} & \$12,281,787 & 0.3\% & 0.4840 & 3.8\% & 0.6732 & 2.3\% & 0.6193 \\
\{0.75,0.6,0.25\} & \$11,150,737 & 0.1\% & 0.3856 & 4.7\% & 0.6736 & 2.4\% & 0.6155  \\
\{0.85,0.5,0.25\} & \$10,907,610 & 0.1\% & 0.3790 & 4.4\% & 0.6850 & 3.1\% & 0.6325  \\
\textbf{Scenario 3} & & & & & & &  \\
\{0.7,0.8,0.95\} & \$17,643,036 & 2.7\% & 0.8341 & 0.2\% & 0.7042 & 0.04\% & 0.6655  \\
\{0.5,0.7,0.95\} & \$15,944,289 & 16.9\% & 0.8798 & 0.5\% & 0.6386 & 0.4\% & 0.5976 \\
\{0.4,0.6,0.85\} & \$13,976,716 & 16.6\% & 0.8419 & 0.7\% & 0.5917 & 0.4\% & 0.5491  \\
\{0.4,0.6,0.75\} & \$13,222,682 & 12.1\% & 0.7759 & 1.2\% & 0.5909 & 0.5\% & 0.5446  \\
\{0.25,0.5,0.85\} & \$12,520,572 & 28.6\% & 0.8943 & 0.4\% & 0.5403 & 0.2\% & 0.4956  \\
\{0.25,0.6,0.75\} & \$12,494,872 & 15.1\% & 0.8056 & 0.5\% & 0.5533 & 0.2\% & 0.5062  \\
\hline
\end{tabular}
}
\caption{Simulation results (R = 10,000) for attribute estimators (SA, SSA and CSA) ($n_1 = 4, n_2 = 5, n_3 = 5, n_4 = 4$).}
\label{tab:sim_results_grouped}
\end{table}
\newpage
\begin{table}[ht]
\centering
{\scriptsize
\setlength{\tabcolsep}{4pt}
\renewcommand{\arraystretch}{1.1}

\begin{tabular}{| l | c | c | c | c | c | c | c | c |}
\hline
 & &
\multicolumn{2}{|c|}{\textbf{Mean}} &
\multicolumn{2}{c|}{\textbf{Difference}} &
\multicolumn{3}{c|}{\textbf{Minimum SE}} \\
\cline{3-4} \cline{5-6} \cline{7-9}
\textbf{\{$p_2,p_3,p_4$\}} &  \textbf{E[$T_Q$]} &
\% LCB Fails & E[$\frac{\text{LCB}_{SM}(\hat{T_Q})}{T_Q}$] &
\% LCB Fails & E[$\frac{\text{LCB}_{SD}(\hat{T_Q})}{T_Q}$] &
\% LCB Fails & E[$\frac{\text{LCB}_{\mathbf{\cdot}}(\hat{T_Q})}{T_Q}$] & \% SD \\
\hline
\textbf{Scenario 1}  & & & & & & & & \\
\{0.95,0.95,0.95\} & \$19,948,553 & 4.2\% & 0.8578 & 28.3\% & 0.9354 & 27.7\% & 0.9376 & 95.8\%\\
\{0.85,0.85,0.85\} & \$17,980,979 & 4.9\% & 0.8182 & 14.1\% & 0.8648 & 13.2\% & 0.8653 & 84.8\% \\
\{0.75,0.75,0.75\} & \$16,013,405 & 4.3\% & 0.7752 & 10.3\% & 0.8063 & 10.9\% & 0.8168 & 74.9\%\\
\{0.65,0.65,0.65\} & \$14,045,832 & 4.2\% & 0.7322 & 8.4\% & 0.7533 & 8.8\% & 0.7676 & 64.3\%\\
\{0.5,0.5,0.5\} & \$11,094,471 & 3.6\% & 0.6626 & 6.8\% & 0.6593  & 7.2\% & 0.6901 & 49.0\%\\
\{0.25,0.25,0.25\} & \$6,175,536 & 1.9\% & 0.4979 & 5.7\% & 0.4422 & 4.9\% & 0.5266 & 25.8\%\\
\textbf{Scenario 2}  & & & & & & & &\\
\{0.95,0.8,0.7\} & \$16,970,969 & 4.1\% & 0.7945 & 11.1\% & 0.8241 & 11.5\% & 0.8361 & 74.1\%\\
\{0.95,0.7,0.5\} & \$14,734,568 & 4.0\% & 0.7603 & 8.1\% & 0.7688  & 8.8\% & 0.7890 & 58.1\%\\
\{0.85,0.6,0.4\} & \$12,766,994 & 3.1\% & 0.7180 & 7.0\% & 0.7178 & 6.7\% & 0.7406 & 47.5\%\\
\{0.75,0.6,0.4\} & \$12,281,787 & 3.3\% & 0.7027 & 7.2\% & 0.7022 & 6.8\% & 0.7294 & 47.0\%\\
\{0.75,0.6,0.25\} & \$11,150,737 & 3.0\% & 0.6994 & 6.4\% & 0.6750 & 6.9\% & 0.7217 & 36.2\%\\
\{0.85,0.5,0.25\} & \$10,907,610 & 2.6\% & 0.6933 & 6.2\% & 0.6683 & 6.4\% & 0.7174 & 35.1\%\\
\textbf{Scenario 3}  & & & & & & & &\\
\{0.7,0.8,0.95\} & \$17,643,036 & 4.5\% & 0.8168 & 11.8\% & 0.8670 & 11.5\% & 0.8678 & 91.8\%\\
\{0.5,0.7,0.95\} & \$15,944,289 & 4.2\% & 0.7922 & 8.8\% & 0.8329  & 8.9\% & 0.8358 & 89.1\%\\
\{0.4,0.6,0.85\} & \$13,976,716 & 4.3\% & 0.7459 & 8.0\% & 0.7765 & 8.8\% & 0.7856 & 77.1\%\\
\{0.4,0.6,0.75\} & \$13,222,682 & 3.9\% & 0.7191 & 8.7\% & 0.7495 & 8.5\% & 0.7595 & 69.3\%\\
\{0.25,0.5,0.85\} & \$12,520,572 & 4.0\% & 0.7171 & 7.7\% & 0.7523 & 7.7\% & 0.7596 & 74.8\%\\
\{0.25,0.6,0.75\} & \$12,494,872 & 3.6\% & 0.7072 & 8.1\% & 0.7327 & 8.0\% & 0.7449 & 68.3\%\\
\hline
\end{tabular}
}
\caption{Simulation results (R = 10,000) for stratified mean and difference estimators (SM, SD) without resampling ($n_1 = 4, n_2 = 9, n_3 = 9, n_4 = 8$).}
\label{tab:sim_results_grouped}
\end{table}
\newpage
\begin{table}[ht]
\centering
{\scriptsize
\setlength{\tabcolsep}{4pt}
\renewcommand{\arraystretch}{1.1}

\begin{tabular}{| l | c | c | c | c | c | c | c | c |}
\hline
 & &
\multicolumn{2}{|c|}{\textbf{Mean}} &
\multicolumn{2}{c|}{\textbf{Difference}} &
\multicolumn{3}{c|}{\textbf{Minimum SE}} \\
\cline{3-4} \cline{5-6} \cline{7-9}
\textbf{\{$p_2,p_3,p_4$\}} &  \textbf{E[$T_Q$]} &
\% LCB Fails & E[$\frac{\text{LCB}_{SM}(\hat{T_Q})}{T_Q}$] &
\% LCB Fails & E[$\frac{\text{LCB}_{SD}(\hat{T_Q})}{T_Q}$] &
\% LCB Fails & E[$\frac{\text{LCB}_{\mathbf{\cdot}}(\hat{T_Q})}{T_Q}$] & \% SD \\
\hline
\textbf{Scenario 1}  & & & & & & & &\\
\{0.95,0.95,0.95\} & \$19,948,553 & 3.2\% & 0.7816 & 48.2\% & 0.9098 & 48.2\% & 0.9165 & 92.8\%\\
\{0.85,0.85,0.85\} & \$17,980,979 & 4.1\% & 0.7203 & 17.4\% & 0.7967 & 17.3\% & 0.8130 & 81.1\%\\
\{0.75,0.75,0.75\} & \$16,013,405 & 4.4\% & 0.6595 & 11.4\% & 0.7083 & 12.0\% & 0.7355 & 70.9\%\\
\{0.65,0.65,0.65\} & \$14,045,832 & 3.6\% & 0.5978 & 9.7\% & 0.6310 & 9.8\% & 0.6712 & 61.9\%\\
\{0.5,0.5,0.5\} & \$11,094,471 & 2.9\% & 0.4922 & 7.8\% & 0.5091 & 7.9\% & 0.5647 & 49.6\%\\
\{0.25,0.25,0.25\} & \$6,175,536 & 1.3\% & 0.2633 & 5.5\% & 0.2875 & 4.5\% & 0.3327 & 28.4\%\\
\textbf{Scenario 2}  & & & & & & & &\\
\{0.95,0.8,0.7\} & \$16,970,969 & 4.4\% & 0.6867 & 13.2\% & 0.7289 & 12.3\% & 0.7596 & 70.4\%\\
\{0.95,0.7,0.5\} & \$14,734,568 & 2.8\% & 0.6328 & 8.7\% & 0.6388  & 9.1\% & 0.6920 & 55.4\%\\
\{0.85,0.6,0.4\} & \$12,766,994 & 2.4\% & 0.5738 & 7.0\% & 0.5675 & 7.4\% & 0.6334 & 48.0\%\\
\{0.75,0.6,0.4\} & \$12,281,787 & 2.8\% & 0.5579 & 6.8\% & 0.5457  & 7.4\% & 0.6139 & 47.0\%\\
\{0.75,0.6,0.25\} & \$11,150,737 & 2.1\% & 0.5445 & 6.6\% & 0.5125\ & 6.6\% & 0.6053 & 37.8\% \\
\{0.85,0.5,0.25\} & \$10,907,610 & 2.4\% & 0.5427 & 6.4\% & 0.5092 & 6.8\% & 0.6016 & 37.7\%\\
\textbf{Scenario 3}  & & & & & & & &\\
\{0.7,0.8,0.95\} & \$17,643,036 & 3.7\% & 0.7244 & 13.2\% & 0.8054 & 13.2\% & 0.8075 & 86.5\%\\
\{0.5,0.7,0.95\} & \$15,944,289 & 3.7\% & 0.6861 & 10.6\% & 0.7582 & 10.6\% & 0.7622 & 82.7\%\\
\{0.4,0.6,0.85\} & \$13,976,716 & 4.1\% & 0.6168 & 9.1\% & 0.6711 & 9.3\% & 0.6932 & 72.7\%\\
\{0.4,0.6,0.75\} & \$13,222,682 & 4.2\% & 0.5856 & 9.5\% & 0.6273 & 9.5\% & 0.6594 & 66.2\%\\
\{0.25,0.5,0.85\} & \$12,520,572 & 3.5\% & 0.5759 & 7.8\% & 0.6369 & 8.5\% & 0.6513 & 70.2\%\\
\{0.25,0.6,0.75\} & \$12,494,872 & 4.0\% & 0.5598 & 8.9\% & 0.6087 & 9.9\% & 0.6397 & 66.0\%\\
\hline
\end{tabular}
}
\caption{Simulation results (R = 10,000) for stratified mean and difference estimators (SM, SD) without resampling ($n_1 = 4, n_2 = 5, n_3 = 5, n_4 = 4$).}
\label{tab:sim_results_grouped}
\end{table}
\newpage
\begin{table}[ht]
\centering
{\scriptsize
\setlength{\tabcolsep}{4pt}
\renewcommand{\arraystretch}{1.1}

\begin{tabular}{| l | c | c | c | c | c | c | c | c |}
\hline
 & &
\multicolumn{2}{|c|}{\textbf{Mean}} &
\multicolumn{2}{c|}{\textbf{Difference}} &
\multicolumn{3}{c|}{\textbf{Minimum SE}} \\
\cline{3-4} \cline{5-6} \cline{7-9}
\textbf{\{$p_2,p_3,p_4$\}} &  \textbf{E[$T_Q$]} &
\% LCB Fails & E[$\frac{\text{LCB}_{SM}(\hat{T_Q})}{T_Q}$] &
\% LCB Fails & E[$\frac{\text{LCB}_{SD}(\hat{T_Q})}{T_Q}$] &
\% LCB Fails & E[$\frac{\text{LCB}_{\mathbf{\cdot}}(\hat{T_Q})}{T_Q}$] & \% SD \\
\hline
\textbf{Scenario 1}  & & & & & & & & \\
\{0.95,0.95,0.95\} & \$19,948,553 & 50.4\% & 0.9935 & 27.6\% & 0.9149 & 27.6\% & 0.9222 & 95.7\%\\
\{0.85,0.85,0.85\} & \$17,980,979 & 33.0\% & 0.9415 & 10.4\% & 0.8071 & 10.7\% & 0.8356 & 85.8\%\\
\{0.75,0.75,0.75\} & \$16,013,405 & 23.9\% & 0.8915 & 5.6\% & 0.7110 & 7.1\% & 0.7745 & 75.8\%\\
\{0.65,0.65,0.65\} & \$14,045,832 & 18.2\% & 0.8386 & 3.3\% & 0.6171 & 5.6\% & 0.7201 & 65.4\%\\
\{0.5,0.5,0.5\} & \$11,094,471 & 11.7\% & 0.7485 & 1.6\% & 0.4392 & 4.0\% & 0.6273 & 50.3\%\\
\{0.25,0.25,0.25\} & \$6,175,536 & 4.5\% & 0.5330 & 0.3\% & 0.0916 & 1.9\% & 0.4340 & 24.3\%\\
\textbf{Scenario 2} & & & & & & & &\\
\{0.95,0.8,0.7\} & \$16,970,969 & 22.5\% & 0.9003 & 6.0\% & 0.7306 & 8.0\% & 0.7930 & 73.8\%\\
\{0.95,0.7,0.5\} & \$14,734,568 & 14.1\% & 0.8404 & 2.1\% & 0.6145 & 4.8\% & 0.7362 & 56.8\%\\
\{0.85,0.6,0.4\} & \$12,766,994 & 10.7\% & 0.7890 & 1.5\% & 0.5146 & 4.0\% & 0.6876 & 46.9\%\\
\{0.75,0.6,0.4\} & \$12,281,787 & 10.7\% & 0.7785 & 1.4\% & 0.4908 & 3.8\% & 0.6714 & 46.4\%\\
\{0.75,0.6,0.25\} & \$11,150,737 & 7.8\% & 0.7526 & 0.6\% & 0.4183 & 3.9\% & 0.6706 & 33.4\%\\
\{0.85,0.5,0.25\} & \$10,907,610 & 7.3\% & 0.7483 & 0.5\% & 0.4031 & 2.9\% & 0.6616 & 33.2\%\\
\textbf{Scenario 3} & & & & & & & & \\
\{0.7,0.8,0.95\} & \$17,643,036 & 37.1\% & 0.9603 & 8.7\% & 0.8320 & 8.9\% & 0.8430 & 94.0\%\\
\{0.5,0.7,0.95\} & \$15,944,289 & 32.7\% & 0.9408 & 6.1\% & 0.7851 & 6.3\% & 0.8027 & 92.3\%\\
\{0.4,0.6,0.85\} & \$13,976,716 & 24.9\% & 0.8838 & 4.1\% & 0.6823 & 5.6\% & 0.7380 & 80.2\%\\
\{0.4,0.6,0.75\} & \$13,222,682 & 20.0\% & 0.8437 & 3.8\% & 0.6201 & 5.5\% & 0.7090 & 70.1\%\\
\{0.25,0.5,0.85\} & \$12,520,572 & 23.3\% & 0.8655 & 3.3\% & 0.6360 & 4.7\% & 0.7074 & 78.2\%\\
\{0.25,0.6,0.75\} & \$12,494,872 & 20.0\% & 0.8347 & 3.6\% & 0.5935 & 5.6\% & 0.6926 & 70.2\%\\
\hline
\end{tabular}
}
\caption{Simulation results (R = 10,000) for stratified mean and difference estimators WITH resampling ($n_1 = 4, n_2 = 9, n_3 = 9, n_4 = 8$).}
\label{tab:sim_results_grouped}
\end{table}
\newpage
\begin{table}[ht]
\centering
{\scriptsize
\setlength{\tabcolsep}{4pt}
\renewcommand{\arraystretch}{1.1}

\begin{tabular}{| l | c | c | c | c | c | c | c | c |}
\hline
 & &
\multicolumn{2}{|c|}{\textbf{Mean}} &
\multicolumn{2}{c|}{\textbf{Difference}} &
\multicolumn{3}{c|}{\textbf{Minimum SE}} \\
\cline{3-4} \cline{5-6} \cline{7-9}
\textbf{\{$p_2,p_3,p_4$\}} &  \textbf{E[$T_Q$]} &
\% LCB Fails & E[$\frac{\text{LCB}_{SM}(\hat{T_Q})}{T_Q}$] &
\% LCB Fails & E[$\frac{\text{LCB}_{SD}(\hat{T_Q})}{T_Q}$] &
\% LCB Fails & E[$\frac{\text{LCB}_{\mathbf{\cdot}}(\hat{T_Q})}{T_Q}$] & \% SD \\
\hline
\textbf{Scenario 1}  & & & & & & & & \\
\{0.95,0.95,0.95\} & \$19,948,553 & 45.6\% & 0.9634 & 49.4\% & 0.8770 & 49.8\% & 0.9016 & 93.5\%\\
\{0.85,0.85,0.85\} & \$17,980,979 & 30.9\% & 0.8766 & 15.2\% & 0.6916 & 16.0\% & 0.7626 &80.8\%\\
\{0.75,0.75,0.75\} & \$16,013,405 & 21.2\% & 0.7860 & 7.9\% & 0.5409 & 9.7\% & 0.6631 & 71.6\%\\
\{0.65,0.65,0.65\} & \$14,045,832 & 14.6\% & 0.6977 & 5.2\% & 0.4091 & 6.5\% & 0.5658 & 62.5\%\\
\{0.5,0.5,0.5\} & \$11,094,471 & 9.3\% & 0.5557 & 2.4\% & 0.2173 & 4.0\% & 0.4290 & 49.3\%\\
\{0.25,0.25,0.25\} & \$6,175,536 & 2.7\% & 0.2465 & 0.5\% & 0.0321 & 1.4\% & 0.1917 & 28.9\%\\
\textbf{Scenario 2} & & & & & & & &\\
\{0.95,0.8,0.7\} & \$16,970,969 & 19.2\% & 0.7960 & 10.3\% & 0.5671 & 10.5\% & 0.6884 & 69.3\%\\
\{0.95,0.7,0.5\} & \$14,734,568 & 10.3\% & 0.7030 & 4.0\% & 0.3815 & 5.3\% & 0.5834 & 54.8\%\\
\{0.85,0.6,0.4\} & \$12,766,994 & 7.0\% & 0.6261 & 2.1\% & 0.2647 & 4.1\% & 0.5170 & 45.9\%\\
\{0.75,0.6,0.4\} & \$12,281,787 & 7.0\% & 0.6030 & 2.1\% & 0.2436 & 3.9\% & 0.4922 & 45.3\%\\
\{0.75,0.6,0.25\} & \$11,150,737 & 4.9\% & 0.5793 & 0.8\% & 0.1673 & 3.3\% & 0.4774 & 34.9\%\\
\{0.85,0.5,0.25\} & \$10,907,610 & 4.6\% & 0.5680 & 0.9\% & 0.1487 & 2.9\% & 0.4671 & 35.5\%\\
\textbf{Scenario 3} & & & & & & & & \\
\{0.7,0.8,0.95\} & \$17,643,036 & 35.4\% & 0.9194 & 11.3\% & 0.7399 & 12.6\% & 0.7775 & 90.4\%\\
\{0.5,0.7,0.95\} & \$15,944,289 & 31.0\% & 0.8911 & 7.3\% & 0.6759 & 8.7\% & 0.7194 & 87.9\%\\
\{0.4,0.6,0.85\} & \$13,976,716 & 24.1\% & 0.7876 & 5.5\% & 0.5288 & 7.2\% & 0.6169 & 76.6\%\\
\{0.4,0.6,0.75\} & \$13,222,682 & 19.0\% & 0.7108 & 5.3\% & 0.4322 & 6.8\% & 0.5580 & 68.0\%\\
\{0.25,0.5,0.85\} & \$12,520,572 & 23.3\% & 0.7565 & 4.0\% & 0.4630 & 7.2\% & 0.5738 & 75.2\%\\
\{0.25,0.6,0.75\} & \$12,494,872 & 19.3\% & 0.7009 & 5.0\% & 0.4044 & 6.7\% & 0.5368 & 66.8\%\\
\hline
\end{tabular}
}
\caption{Simulation results (R = 10,000) for stratified mean and difference estimators WITH resampling ($n_1 = 4, n_2 = 5, n_3 = 5, n_4 = 4$).}
\label{tab:sim_results_grouped}
\end{table}

%\pagenumbering{empty}      % normal numbers starting with 1

\end{document}